\documentclass[letterpaper, 10 pt, conference]{ieeeconf}  % Comment this line out if you need a4paper

\IEEEoverridecommandlockouts                              % This command is only needed if 
\usepackage{amsmath} % assumes amsmath package installed
\usepackage{amssymb}  % assumes amsmath package installed
\usepackage{hyperref}
\usepackage{dsfont}
\usepackage{xcolor}
\usepackage{algorithm, algpseudocode}
\usepackage{mathtools}
\usepackage{enumerate}

\usepackage{tikz,pgfplots}

\usetikzlibrary{arrows.meta}

\usepackage{multirow}
\usepackage{makecell}
\usepackage{adjustbox}
\usepackage{optidef}
\usepackage[compress]{cite}

\pgfplotsset{compat=1.17} 
\usetikzlibrary{pgfplots.groupplots}

\newtheorem{assumption}{Assumption}
\newtheorem{standing assumption}{Standing Assumption}

\newtheorem{theorem}{Theorem}

\newtheorem{corollary}{Corollary}
\newtheorem{definition}{Definition}

\newtheorem{remark}{Remark}

\newcommand{\mc}{\mathcal}
\newcommand{\til}{\Tilde}
\newcommand{\ov}{\overline}
\newcommand{\ul}{\underline}
\newcommand{\Z}{\mathbb{Z}}
\newcommand{\Prob}{\mathbb{P}}

\newcommand{\R}{\mathds{R}}

\newcommand{\tcr}{\textcolor{red}}

\newcommand{\defineas}{\coloneqq}

\newcommand{\E}{\mathbb{E}}

\newcommand{\norm}[1]{\left\lVert#1\right\rVert}

\title{\LARGE \bf
No-Regret Learning in Stackelberg Games with an Application to Electric Ride-Hailing
}

\author{Anna Maddux$^*$, Marko Maljkovic$^*$, Nikolas Geroliminis, and Maryam Kamgarpour% <-this % stops a space
\thanks{Anna Maddux and Maryam Kamgarpour are with SYCAMORE group, Institute of Mechanical Engineering, EPFL Switzerland. Marko Maljkovic and Nikolas Geroliminis are with LUTS group, School of Architecture, Civil and Environmental Engineering, EPFL Switzerland. \tt\small \{anna.maddux, marko.maljkovic, maryam.kamgarpour, nikolas.geroliminis\}@epfl.ch}%
\thanks{This work was supported by the Swiss National Science Foundation under NCCR Automation Phase 2 project, fonds 565 776 and NCCR Automation Phase 1, fonds 565422.}%
\thanks{*The first two authors contributed equally to this work.}
}%

\begin{document}

\maketitle
\thispagestyle{empty}
\pagestyle{empty}

%%%%%%%%%%%%%%%%%%%%%%%%%%%%%%%%%%%%%%%%%%%%%%%%%%%%%%%%%%%%%%%%%%%%%%%%%%%%%%%%
\begin{abstract}
We consider the problem of efficiently learning to play single-leader multi-follower Stackelberg games when the leader lacks knowledge of the lower-level game. Such games arise in hierarchical decision-making problems involving self-interested agents. For example, in electric ride-hailing markets, a central authority aims to learn optimal charging prices to shape fleet distributions and charging patterns of ride-hailing companies. Existing works typically apply gradient-based methods to find the leader's optimal strategy. Such methods are impractical as they require that the followers share private utility information with the leader. Instead, we treat the lower-level game as a black box, assuming only that the followers' interactions approximate a Nash equilibrium while the leader observes the realized cost of the resulting approximation. Under kernel-based regularity assumptions on the leader's cost function, we develop a no-regret algorithm that converges to an $\epsilon$-Stackelberg equilibrium in $\mc O(\sqrt{T})$ rounds. Finally, we validate our approach through a numerical case study on optimal pricing in electric ride-hailing markets. 

\end{abstract}

\section{Introduction}

Many real-world systems, including traffic networks, communication systems, and smart grids, involve multiple self-interested agents that interact repeatedly. Such independent and self-interested decision-making may result in inefficient and socially undesirable outcomes due to misaligned objectives and a lack of coordination among the agents \cite{dubey1986}. Therefore, from a global perspective, it is important to design intervention mechanisms such as tolls, pricing schemes, and subsidies that can steer agents towards a socially desirable and efficient behavior.

A prominent example of inefficient decision-making arises in ride-hailing markets, where service providers may flock to regions of high demand in search of passengers, leaving other customer locations underserved~\cite{BEOJONE2021102890}. Interestingly, the growing integration of electric vehicles (EVs) into ride-hailing fleets of companies such as Uber, Lyft, and Curb, introduces a lever for external intervention via electricity pricing \cite{maljkovic2023hierarchical,Li2023taxi, malkovic2023learninghow}. A central authority, such as the government or power providers, may set spatially varying prices to shape fleet distribution and charging patterns to match demand and supply across regions.

Introducing a central authority to design intervention mechanisms imposes a bi-level structure that naturally fits into the framework of a single-leader, multi-follower Stackelberg game~\cite{von1952theory}. In this setting, the incentive designer assumes the role of the leader, while the self-interested agents act as followers. In the context of ride-hailing, the government or the power provider would act as the leader, while the competing ride-hailing fleets would serve as the followers. At the upper level, the leader selects incentives to minimize a societal cost, anticipating the followers' best responses. At the lower level, the followers are considered rational agents that play a non-cooperative game, the outcome of which forms a Nash equilibrium. %The main question we address in this paper is whether Stackelberg games can be efficiently solved when the central authority must account for the lower-level game's behavioral nature in addressing the incentive design problem described above.
This paper investigates whether the leader can efficiently learn 
to solve Stackelberg games solely from observations of the societal cost to tackle the incentive design problem described above. % to respect the behavioral nature of the lower-level game. %in addressing the incentive design problem described above.

%A solution to the MPEC is a Stackelberg equilibrium, which is in general hard to compute due to ill-posed nature of the problem. Namely, MPECs are often non-convex, non-smooth, violate standard constraint qualifications, and exhibit irregular or even disconnected feasible regions~\cite{luo1996mathematical, colson2007overview}. 

Existing approaches usually rely on first-order methods~\cite{grontas24, maljkovic2024decentralizedcomputationleadersstrategy, liu2022inducing, ji2021bilevel,goyal2023algorithm} or problem-specific solutions~\cite{l-se}. 
% more references \cite{fiez2019convergence, thoma2024contextual}
First-order methods typically estimate the gradient of the leader's objective, i.e., the so-called hypergradient, but this estimation requires the leader to have access to estimates of both the lower-level Nash equilibrium and the gradient of the followers' utility functions \cite{grontas24}. While these methods may be suitable for bilevel optimization problems \cite{anandalingam1992hierarchical}, they become impractical when the lower-level involves self-interested agents that are unwilling to share their private information. In particular, this is evident in electric ride-hailing markets, where agents behave strategically and are unlikely to disclose sensitive information, such as the gradient of their utility functions. As a result, the lower-level problem effectively becomes a black box to the leader, making traditional approaches difficult to apply.
Several past works \cite{ sessa2020learning,maheshwari2023follower, zhao2023online} allow the leader to probe the followers with different incentives and observe the realized societal cost. The work of \cite{sessa2020learning} focuses on a single-leader, single-follower setup, where the leader observes the follower's noisy best response. Assuming that the follower's response function lies in a reproducing kernel Hilbert space (RKHS), they propose a no-regret algorithm that converges to an approximately optimal incentive in $\mc O(\sqrt{T})$. In contrast, \cite{maheshwari2023follower} considers a different approach to tackle multi-follower Stackelberg games based on zeroth-order estimation of the leader's hypergradient. While their algorithm converges to an approximately stationary point in $\mc O(\sqrt{T})$, to ensure that it is an approximate Stackelberg equilibrium, they require a hard-to-verify assumption on the Hessian of the leader's cost function. Moreover, their method estimates the hypergradient by probing followers with an incentive and a perturbed version thereof, which may be impractical when a central authority cannot evaluate multiple incentives simultaneously.

In this paper, we consider a class of multi-follower Stackelberg games
%for which we propose a provably efficient algorithm based on one-point feedback, which as such is compatible with the behavioral assumptions of a game. 
%We summarize our contributions as follows:
for which our contributions are threefold:
% \tcr{Update, also I think we should check the relevance of \cite{cesa2023learning,goyal2023algorithm}, also comment on incentive design problem}
\begin{itemize}
    \item We propose a novel no-regret algorithm for the leader to learn in Stackelberg games that leverages Gaussian process regression, assuming that the leader's cost function satisfies the RKHS assumption.
    %We propose a novel no-regret algorithm for the leader, called \underline{B}est \underline{I}ntervention in \underline{G}ames using \underline{G}aussian \underline{P}rocesses and \underline{L}ower \underline{C}onfidence \underline{B}ounds (BIG GP-LCB), for playing Stackelberg games when the leader's cost function satisfies the RKHS assumption. 
    \item We show that with high probability our algorithm converges in $\mc O(\sqrt{T})$ to an $\epsilon$-Stackelberg equilibrium. Our method requires no prior knowledge of the lower-level game and only assumes that the followers can approximate a Nash equilibrium within a number of rounds polynomial in $T$. 
    \item We demonstrate the applicability of our setup to electric ride-hailing markets and validate our algorithm through a numerical case study in this domain.
\end{itemize}

%Compared to \cite{sessa2020learning}, our multi-follower setting is more challenging because the followers only find an approximate Nash equilibrium and can not find their exact best response. Compared to \cite{maheshwari2023follower}, our assumptions on the leader's cost function are less stringent, and our algorithm requires only one-point feedback.

%Compared to [14], our multi-follower setting introduces additional complexity, as followers only reach an approximate Nash equilibrium rather than their exact best response. Relative to [15], our assumptions on the leader’s cost function are less restrictive, and our approach requires only one-point feedback.

% The paper is outlined as follows: the rest of this section introduces the basic notation. In Sections~\ref{sec:problem_setup} and~\ref{sec:mot_examp}, we introduce the general Stackelberg problem and reformulate it in the context of our motivating example on coordination in the electric ride-hailing markets. Our main methodological and theoretical results are discussed in Section~\ref{sec:algorithm}, followed by numerical results in Section~\ref{sec:numerical}. Finally, Section~\ref{sec:conclusion} concludes the paper with remarks and suggestions for future research.

\noindent\textit{Notation:} Let $\R$ and $\Z_{(+)}$ denote the sets of real and (non-negative) integer numbers. For any $T\in\Z_+$, we let $[T]=\{1,2,...,T\}$. Let $\mathds{1}$ be the vector of all ones. If $\mc A$ is a finite index, we let $x=(x_i)_{i\in\mc A}$ be the concatenation of vectors $x_i$. For a real-valued function $f:\R^d\rightarrow\R$, we let $\nabla_{x}f=(\frac{\partial f}{\partial x_i})_{i=1}^d$ be its gradient.

\section{Problem setup}\label{sec:problem_setup}

We consider a Stackelberg game with $N+1$ agents consisting of a leader $L$ and a set of followers denoted by $\mc N$. In a Stackelberg game, the leader first chooses an action $\pi\in \Pi$ from its action set $\Pi\subseteq \R^p$. The followers then simultaneously respond to the leader's action $\pi$ by choosing an action $x_i\in\mc X_i\subseteq \R^d$, where $\mc X_i$ denotes the action set of follower $i\in\mc N$.
Furthermore, the leader has a cost function given by $J:\Pi \times \mc X\rightarrow\R$ and each follower has a utility function given by $U_i:\mc X \times \Pi\rightarrow\R$, where $\mc X=\Pi_{i=1}^N\mc X_i$ denotes the action space of the joint action profile $x\defineas(x_i)_{i\in\mc N}$.

\subsection{Lower-level game}
Given the leader's action $\pi$, at the lower level, the followers decision-making problem can be cast as a game $\Gamma(\mc N, \{\mc X_i\}_{i=1}^N, \{U_i\}_{i=1}^N; \pi)$. Each agent aims to maximize its utility given the actions of the other agents and the action of the leader. A popular solution concept in such games is the Nash equilibrium at which no agent has an incentive to unilaterally deviate from its action.

% \begin{definition}
%     Given $\pi$, the joint action profile $x^*\in\mc X$ is an $\epsilon$-Nash equilibrium, if the following holds:
%     \begin{align*}
%         U_i(x_i,x_{-i};\pi)\leq U_i(x_i^*,x_{-i}^*;\pi)+\epsilon,\quad\forall x_i\in\mc X_i, \forall i\in\mc N,
%     \end{align*}
%     where $x_{-i}\coloneqq(x_j)_{j\in\mc N\setminus\{i\}}$ is the joint action of all agents except $i$.
%     The action profile $x^*$ is a Nash equilibrium if $\epsilon=0$.
% \end{definition}
\begin{definition}
    For a given $\pi$, the joint action profile $x^*\in\mc X$ is an $\epsilon$-Nash equilibrium if $U_i(x_i,x_{-i}^*;\pi)\leq U_i(x_i^*,x_{-i}^*;\pi)+\epsilon$ holds for all $i\in\mc N$, where $x_{-i}\coloneqq(x_j)_{j\in\mc N\setminus\{i\}}$ denotes the joint action of all agents except $i$. The action profile $x^*$ is a Nash equilibrium if $\epsilon=0$.
\end{definition}

In the following, we assume that the game $\Gamma$ is concave and strongly monotone. Concavity ensures that a Nash equilibrium exists \cite[Theorem~1]{rosen1965existence}, while strong monotonicity ensures that it is also unique \cite[Theorem~2]{rosen1965existence}. 
% could also cite \cite{debreu1952social} for existence

\begin{assumption}\label{ass:concave_game}
    $\Gamma(\mc N, \{\mc X_i\}_{i=1}^N, \{U_i\}_{i=1}^N; \pi)$ is a concave game for every $\pi\in\Pi$. Namely, for each agent $i\in\mc N$ the set $\mc X_i$ is non-empty, compact, and convex, and the utility function $U_i(x_i,x_{-i};\pi)$ is continuously differentiable in $x$, and concave in $x_i$ for all $x_{-i}\in \Pi_{j\neq i}\mc X_j$ and all $\pi\in\Pi$.
\end{assumption}
    
\begin{assumption}\label{ass:strongly_monotone_game}
    $\Gamma(\mc N, \{\mc X_i\}_{i=1}^N, \{U_i\}_{i=1}^N; \pi)$ is $\alpha$-strongly monotone for every $\pi\in\Pi$. Namely, the game pseudogradient $v:\R^{Nd}\times\Pi\rightarrow \R^{Nd}$ defined as: 
\begin{align}\label{eq:utility_gradient}
    &v(x;\pi)=(v_i(x;\pi))_{i=1}^N,\ \text{where}\nonumber\\
    &v_i(x;\pi)=-\nabla_{x_i}U_i(x_i,x_{-i};\pi), \quad \forall x\in\mc X, \forall i\in\mc N,\nonumber
\end{align} 
\noindent satisfies $
            \langle v(x;\pi)-v(x';\pi), x-x'\rangle\geq \alpha \norm{x-x'}^2$ for all $x,x'\in\mc X$ and $\alpha>0$.
\end{assumption}

\subsection{Stackelberg game}

Under Assumptions \ref{ass:concave_game} and \ref{ass:strongly_monotone_game}, a Stackelberg game can be expressed as follows:
    \begin{mini!}[left]
        {\pi\in\Pi}{J(\pi,x^*(\pi))}
        {\label{prob:bilevel_optimization}}{\tag{1}}
        \addConstraint{ x_i^*(\pi)\in\arg\max_{x_i\in\mc X_i} U_i(x_i,x_{-i}^*(\pi);\pi), \:\forall i\in\mc N.}{}{}\nonumber
    \end{mini!}
At the upper level, the leader aims to minimize its cost $J$, given the followers' response $x^*(\pi)$. At the lower level, the followers aim to maximize their utility based on the leader's action $\pi$ and the best response of other agents $x_{-i}^*(\pi)$. Under Assumptions \ref{ass:concave_game} and \ref{ass:strongly_monotone_game}, the followers' action profile $x^*(\pi)$ is the unique Nash equilibrium of $\Gamma(\mc N, \{\mc X_i\}_{i=1}^N, \{U_i\}_{i=1}^N; \pi)$, ensuring the Stackelberg game is well-defined since the leader does not have to consider set-valued best reponses.

\iffalse
\begin{align*}
    &\min_{\pi\in\Pi, x^*\in\mc X}\quad J(\pi,x^*)\\
     &s.t.\quad x^*\in S(\pi)\coloneqq SOL(\mc X, v(\cdot;\pi)),
\end{align*}
where $v:\R^{Nd}\times\Pi\rightarrow \R^{Nd}$ is the game pseudogradient defined as
\begin{align*}
    &v(x,\pi)=(-\nabla_{x_i}U_i(x_i,x_{-i};\pi))_{i=1}^N,\ \text{where}\\
    &\nabla_{x_i}U_i(x_i,x_{-i};\pi) = \left(\frac{\partial U_i(x;\pi)}{\partial x_i^k}\right)_{j=1}^d, \quad \forall x\in\mc X, \forall i\in\mc N.
\end{align*}
Furthermore, $SOL(\mc X, v(\cdot;\pi))$ is the solution set to the variational inequality problem $VI(\mc X, v(\cdot;\pi))$ defined as
\begin{align*}
    SOL(\mc X, v(\cdot;\pi))=\{x^*\ \lvert \ \langle v(x^*;\pi), x-x^*\rangle\geq 0\quad\forall x\in\mc X\}.
\end{align*}

$VI(\mc X v(\cdot;\pi))$ is a variational inequality problem, i.e., $\langle v(x^*;\pi), x-x^*\rangle\geq 0$ for all $x\in\mc X$ and $v(x;\pi)=(v_1(x;\pi),\ldots,v_N(x;\pi))$
The map defined by the vector of individual partial gradients:
        \begin{align*}
            v(x;\pi)=(v_1(x;\pi),\ldots,v_N(x;\pi)),
        \end{align*}
        where $v_i(x;\pi)=\nabla_i U_i(x_i,x_{-i};\pi)$
\fi

A stable outcome of a Stackelberg game is the so-called Stackelberg equilibrium, where neither the leader nor the followers have an incentive to unilaterally deviate from their action. We define it as follows:

% \begin{definition}
%     A joint action profile $(\pi^*,x^*)\in\Pi\times\mc X$ is an $\epsilon$-Stackelberg equilibrium if $x^*$ is an $\epsilon$-Nash equilibrium of $\Gamma(\mc N, \{\mc X_i\}_{i=1}^N, \{U_i\}_{i=1}^N; \pi^*)$, i.e., 
%     \begin{align*}
%          U_i(x_i,x_{-i}^*;\pi^*)\leq U_i(x_i^*,x_{-i}^*;\pi^*)+\epsilon,\quad\forall i\in\mc N,
%     \end{align*}
%     and if for all $(\overline{\pi},\overline{x})\in\Pi\times\mc X$ such that $\overline{x}$ is an $\epsilon$-Nash equilibrium of $\Gamma(\mc N, \{\mc X_i\}_{i=1}^N, \{U_i\}_{i=1}^N; \overline{\pi})$ it holds that:
%     \begin{align*}
%         J(\pi^*,x^*)-\epsilon\leq J(\overline{\pi},\overline{x}).
%     \end{align*}
%     The joint action profile $(\pi^*,x^*)$ is a Stackelberg equilibrium if $\epsilon=0$.
% \end{definition}
\begin{definition}
    A joint action profile $(\pi^*,x^*)\in\Pi\times\mc X$ is an $\epsilon$-Stackelberg equilibrium if $x^*$ is an $\epsilon$-Nash equilibrium of $\Gamma(\mc N, \{\mc X_i\}_{i=1}^N, \{U_i\}_{i=1}^N; \pi^*)$, and if for all $(\overline{\pi},\overline{x})\in\Pi\times\mc X$ such that $\overline{x}$ is an $\epsilon$-Nash equilibrium of $\Gamma(\mc N, \{\mc X_i\}_{i=1}^N, \{U_i\}_{i=1}^N; \overline{\pi})$ it holds that $J(\pi^*,x^*)-\epsilon\leq J(\overline{\pi},\overline{x})$.
    The joint action profile $(\pi^*,x^*)$ is a Stackelberg equilibrium if $\epsilon=0$.
\end{definition}

In general, the leader lacks access to the followers' utility functions and, consequently, does not have a closed-form expression for its cost function, making direct computation of a Stackelberg equilibrium impossible. Instead, this paper aims to learn a Stackelberg equilibrium from observations of how the followers respond to the leader’s actions.
Before presenting our method, we first introduce a motivating example from the domain of future smart mobility.

\section{Motivating example: Ride-hailing markets}\label{sec:mot_examp}

We formulate charging management in ride-hailing markets as a Stackelberg game. At the lower level, each ride-hailing company $i\in\mc N$ operates a fleet of $M_i\in\Z$ electric vehicles (EVs) to serve demand across multiple districts, each equipped with its own charging infrastructure. For a given pricing vector $\pi\in\R^d$, companies aim to maximize their profit by optimally dispatching and later recharging their vehicles~\cite{maljkovic2023hierarchical}. Specifically, each company $i$ decides how many EVs to allocate to each of the $d$ districts, choosing $$x_i\in[0,x_{i,\max}]^d\cap\{x_i\in\R^d\ \lvert \ \mathds{1}^\top x_i\leq M_i\}.$$

Given the price vector $\pi$, we assume company $i$'s utility function follows the market share acquisition as in~\cite{10591280, bravo2018bandit}:
\begin{align}\label{eq:Player_util}
        U_i(x_i,x_{-i};\pi) =\sum_{m=1}^d W_m\frac{x_{i,m}}{\sum_{j\in\mc N}x_{j,m}+\Delta_m}-x_{i,m}\pi_m. 
\end{align}
The utility reflects the difference between the company's revenue from serving demand across $d$ districts and recharging costs, where $0< W_{\min}\leq W_m\leq W_{\max}$ represents the total revenue potential of district $m\in[d]$. In high-demand regions, a larger number of service vehicles is required to meet the increased volume of requests and to prevent customer abandonment due to long waiting times. This is captured by the exogenous parameter $0< \Delta_{\min}\leq\Delta_m\leq \Delta_{\max}$, which accounts for the fraction of total revenue potential not being distributed among the ride-hailing companies. For utilities given by~\eqref{eq:Player_util}, it can be verified that the game's pseudogradient is strongly monotone~\cite[Appendix A.2.]{bravo2018bandit}.

At the upper level, the central authority seeks to guide the overall EV allocation towards a predefined target distribution, with the goals of reducing congestion, ensuring equitable service coverage across the city, and balancing grid demand. For instance, charging prices in remote areas may be set lower than in the central districts, encouraging ride-hailing companies to charge their EVs there. Formally speaking, since the central authority lacks information about the absolute number of operating EVs, we assume it sets charging prices $\pi\in[0,\pi_{\max}]^d$ to incentivize ride-hailing companies to match a target distribution $\xi^* \in \{\xi \in [0,1]^d \mid \mathds{1}^\top \xi = 1\}$. The interaction between the companies results in a joint allocation of EVs $x(\pi)=(x_1(\pi),\ldots,x_N(\pi))$, allowing the central authority's cost to be expressed as: 
\begin{align}\label{eq:J_leader}
    J(\pi) = \biggl\lVert \frac{\sum_{i\in\mc N}x_i(\pi)}{\mathds{1}^\top\sum_{i\in\mc N} x(\pi)} - \xi^* \biggr\lVert^2,
\end{align}
which quantifies the deviation between the achieved distribution of EVs and the desired one.
% Each company aims to maximize its utility, the outcome of which is the unique Nash equilibrium $x(\pi)=(x_1(\pi),\ldots,x_N(\pi))$. Then, the cost function of the central authority depends on the joint distribution $x(\pi)$ and is given by:
% \begin{align}\label{eq:J_leader}
%     J(\pi) = \biggl\lVert \frac{x(\pi)}{\mathds{1}^\top x(\pi)} - \xi^* \biggr\lVert^2,
% \end{align}
% which measures the deviation between the actual distribution of EVs resulting from the imposed charging prices and the central authority's desired distribution. 

In reality, the central authority cannot compute the unique Nash equilibrium $x(\pi)$ without knowledge of the companies' utility functions. Instead, it can set charging prices and wait while the companies repeatedly compete with each other for demand until the market stabilizes at an approximation of the unique Nash equilibrium. The central authority can then assess how closely the actual distribution of EVs aligns with its target distribution and adjust the prices accordingly. In the following, we present an algorithm that lets the leader learn an $\epsilon%
$-Stackelberg equilibrium, which is also applicable to our motivating example.

% For the setting of charging management in ride-hailing markets presented in this section, it can be verified that Assumptions \ref{ass:concave_game} - \ref{ass:Lipschitz_cost} hold~\cite[Appendix A.2]{bravo2018bandit}.

\section{Learning a Stackelberg equilibrium}\label{sec:algorithm}

We consider the setting where the Stackelberg game is repeated over several rounds. In each round $t$, the leader selects an action $\pi^t$ that is observed by the followers. Then, in a subroutine, the followers aim to learn an approximation $x^t(\pi^t)$ of the unique Nash equilibrium $x^*(\pi^t)$. %The leader then updates its action $\pi^{t+1}$ based on its observed cost $J(\pi^t,x^t(\pi^t))$.
In the following, we treat the lower-level game as a black-box and merely assume that after finitely many iterations within a subroutine, $x^t(\pi^t)$ is indeed an approximation of $x^*(\pi^t)$. %This can be achieved by several learning algorithms \cite{drusvyatskiy2022improved, bravo2018bandit,tatarenko2022rate}.

We now focus on the leader who seeks to learn a Stackelberg equilibrium in a repeated fashion. As $\pi^t$ changes across rounds, the leader observes a sequence of time-varying costs. This motivates us to define the following notion of regret:
\begin{align}
   R^T =  \sum_{t=1}^T J(\pi^t, x^t(\pi^t))- \min_{\pi\in\Pi}J(\pi, x^*(\pi)).
\end{align}
It measures the leader's additional cost beyond its optimal value due to 1) not knowing $\pi^*\in\arg\min_{\pi\in\Pi} J(\pi,x^*(\pi))$  beforehand and 2) the followers learning an approximation $x^t(\pi^t)$ of the unique Nash equilibrium $x^*(\pi^t)$ given $\pi^t$. The leader has no-regret if $R^T/T\rightarrow 0$ as $T\rightarrow\infty$. As we will show in Theorem \ref{th:1}, presented later in this section, having no-regret implies convergence to a Stackelberg equilibrium.

To ensure that the leader is able to attain no-regret, we make the following assumptions. At the end of round $t$, the leader receives feedback information that it can use to update its action $\pi^{t+1}$ and thus improve its cost. A realistic feedback model is bandit feedback~\cite{Srinivas2010}, where the leader merely observes its realized cost. %In addition, we assume the leader observes the response of the followers. 
%This is in particular true for the motivating example introduced in Section~\ref{sec:mot_examp}.
In ride-hailing applications, this is satisfied since the central authority can assess how closely the actual distribution of EVs matches the target distribution.

\begin{assumption}\label{ass:feedback}
    In round $t$, the leader observes the realized cost $J(\pi^t,x^t(\pi^t))= J(\pi^t, x^*(\pi^t))-\epsilon^t$, where $\epsilon^t$ is the error due to the followers playing an approximation of the unique Nash equilibrium $x^*(\pi^t)$.
\end{assumption}

% regularity assumption
To ensure that the error $\epsilon^t$ is bounded, we assume that the leader's cost function is Lipschitz-continuous with respect to the followers' joint action. 
\begin{assumption}\label{ass:Lipschitz_cost}
    The cost function $J:\Pi\times \mc X\rightarrow \R$ is $L_J$-Lipschitz continuous in $x\in\mc X$, i.e., $\lvert J(\pi,x) - J(\pi,\overline{x})\rvert\leq L_J\lVert x - \overline{x} \rVert$ holds for all $x,\ov x\in\mc X$ and $\pi\in\Pi$.
\end{assumption}

Since $\mc X$ is compact, we can define $M=\max_{x,\ov x\in\mc X}\lVert x-\ov x\rVert$. Then, Assumption \ref{ass:Lipschitz_cost} implies that $\lvert J(\pi,x) - J(\pi,\overline{x})\rvert\leq L_JM$ for all $x,\ov x\in\mc X$ and $\pi\in\Pi$. 

Attaining no-regret is impossible
without any regularity assumptions on the cost function \cite{Srinivas2010}. In this work, we further assume that similar inputs lead to similar outputs. This is satisfied, for example, in ride-hailing applications, where similar charging prices set by the central authority lead to similar vehicle distribution patterns, implying similar costs for the central authority.

\begin{assumption}\label{ass:cost_in_RKHS}
    The cost function {\small$J:\Pi\rightarrow \R$} has a compact domain $\Pi$.\footnote{Note that at the Nash equilibrium $x^*(\pi)$ the leader's cost $J$ is uniquely defined by $\pi$. Thus, with a slight abuse of notation, we can adopt a modified mapping $J:\Pi\rightarrow\R$ with $J(\pi)=J(\pi,x^*(\pi))$.} Furthermore, $J$ has a bounded norm {\small$\|J\|_{k}=\sqrt{\langle J,J\rangle_{k}}\leq B$} in a reproducing kernel Hilbert space (RKHS, see \cite{rasmussen2005}) associated with a positive semi-definite kernel function $k(\cdot,\cdot)$. The RKHS is denoted by {\small$\mc H_{k}(\Pi)$}. We further assume bounded variance by restricting $k(\pi, \pi')\leq 1$ for all $\pi,\pi'\in\Pi$. 
\end{assumption}

This assumption is common in black-box optimization \cite{Srinivas2010} and in repeated games \cite{sessa2020learning, maddux2024multi}. Combined with Assumptions \ref{ass:feedback} and \ref{ass:Lipschitz_cost} when $\epsilon^t$ is $R$-sub-Gaussian, it allows the leader to learn its unknown cost $J$ using the Gaussian process (GP) framework. 

Functions {\small$J\in\mathcal{H}_{k}(\Pi)$} with {\small$\|J\|_{k}<\infty$} can be modeled as a sample from a GP \cite[Section~6.2]{rasmussen2005}, i.e., {\small$J(\cdot) \sim \mathcal{GP}(\mu(\cdot), k(\cdot, \cdot))$}, specified by its mean and covariance functions $\mu(\cdot)$ and $k(\cdot, \cdot)$, respectively. A GP prior {\small$\mathcal{GP}(0,k(\cdot,\cdot))$} over the initial distribution of the unknown function $J$, is used to capture the uncertainty over $J$. Then, for any $\pi\in\Pi$, the function value $J(\pi)$ can be predicated based on a history of measurements $\{y^{\tau}\}_{\tau=1}^t$ at points $\{\pi^{\tau}\}_{\tau=1}^t$, with {\small$y^\tau = J(\pi^{\tau},x^*(\pi^{\tau}))+\epsilon^\tau$} and {\small$\epsilon^\tau\sim\mathcal{N}(0,\sigma^2)$}. Conditioned on the history of measurements, the posterior distribution over $J$ is a GP with mean and variance functions:
{
\begin{align}
    &\mu^t(\pi) = \textbf{k}^t(\pi)^\top (\textbf{K}^t+\sigma^2 \textbf{I}^t)^{-1}\textbf{y}^t \label{eq:mu}\\   
    &(\sigma^t)^2(\pi)=k(\pi,\pi)-\textbf{k}^t(\pi)^\top(\textbf{K}^t+\sigma^2 \textbf{I})^{-1}\textbf{k}^t(\pi), \label{eq:var}
\end{align}
}where $\textbf{k}^t(\pi)=(k(\pi^{\tau},\pi))_{\tau=1}^t$, $\textbf{y}^t=(y^\tau)_{\tau=1}^t$, and $\textbf{K}^t=(k(\pi^{\tau},\pi^{\tau'}))_{\tau,\tau'=1}^t$ is the kernel matrix.

In our setting, however, the leader observes $J(\pi^t,x^t(\pi^t))$ rather than $y^t$. Thus, we define $\til \mu^t(\cdot)$ which also depends on the true observation as follows:
\begin{align}\label{eq:mu_corrected}
\til\mu^t(\pi)=\textbf{k}^t(\pi)^\top(\textbf{K}^t+\sigma^2\textbf{I}^t)^{-1}\mathbf{\til y}^t,
\end{align}
    where $\mathbf{\til y}^t=(J(\pi^\tau,x^\tau(\pi^\tau)))_{\tau=1}^t$.

\iffalse
\tcr{Anna to: add later depending on how the next section develops. Maybe we actually have motivating example next. In the next section, we present our algorithm which exploits the feedback model and the regularity assumption to derive a no-regret no-violation algorithm. We
focus on the perspective of a single agent and drop
the subscript i wherever this is possible without causing confusion.}
\fi

\subsection{Algorithm and analysis}

Next, we present our two-loop algorithm summarized in Algorithm \ref{alg:bilevel_game}. In the outer
loop, the leader selects $\pi^t$ and announces it to the followers. Then, the inner loop, denoted by \textit{ApproxNE$(\pi^t, K)$} in Line 4, runs for $K$ iterations, allowing the followers to learn an approximation of the Nash equilibrium before the leader updates its action again. The inner loop subroutine can be substituted by any learning algorithm that converges to the unique Nash equilibrium of the lower-level game $\Gamma(\mc N, \{\mc X_i\}_{i=1}^N, \{U_i\}_{i=1}^N; \pi^t)$ after finitely many iterations. Formaly speaking, we require that the output of the inner loop $\Tilde{x}(\pi^t)$ satisfies: $$\E[\lVert x^*(\pi^t)-\Tilde{x}(\pi^t)\rVert^2]\leq C(\pi^t)K^{-c}$$ for some constants $C(\pi^t),c\in\R_+$. For instance, the works of~\cite{drusvyatskiy2022improved, bravo2018bandit} propose such methods that satisfy the above condition under Assumptions~\ref{ass:concave_game} and~\ref{ass:strongly_monotone_game}.

Upon observing its realized cost $J(\pi^t,x^t(\pi^t))$, the leader updates its action by choosing the minimizer of the following surrogate cost function:
\begin{align*}
    \ul J^{t+1}(\pi)\coloneqq \til\mu^t(\pi)-\left(\beta^{t+1}+\frac{\epsilon\sqrt{t+1}}{\sigma}\right)\sigma^t(\pi)
\end{align*}
where $\til \mu^t(\cdot)$ and $\sigma^t(\cdot)$ are computed as in Equations \eqref{eq:mu_corrected} and \eqref{eq:var} using past game data {\small$\{(\pi^\tau, J(\pi^\tau,x^\tau(\pi^\tau)))\}_{\tau=1}^{t-1}$}. Parameter $\beta^t$ controls the width of the confidence bound, and if specified adequately, can ensure that $\ul J^{t+1}(\pi)$ represents a lower bounds on $J(\pi)$ that holds with high probability.
\iffalse
\tcr{
\begin{remark}
    Under Assumption \ref{ass:feedback} and \ref{ass:cost_in_RKHS} the leader can learn about its cost function $J$ without needing to observe the followers' joint utility gradient $v(x)$ and without needing to observe the followers' joint action profile $x^t(\pi^t)$.
\end{remark}
}

\fi
% \begin{remark}
%     The inner loop of our algorithm, denoted by \textit{ApproxNE$(\pi^t, K)$} in Line $4$ of Algorithm \ref{alg:bilevel_game}, can be substituted by any learning algorithm that after finitely many iterations within a subroutine converges to the unique Nash equilibrium of the lower-level game $\Gamma(\mc N, \{\mc X_i\}_{i=1}^N, \{U_i\}_{i=1}^N; \pi^t)$. This can be achieved by several learning algorithms \cite{drusvyatskiy2022improved, bravo2018bandit}. For example, Bravo et al. \cite{bravo2018bandit} propose a mirror descent algorithm and establish convergence rate guarantees to the unique Nash equilibrium for games adhering to Assumptions~\ref{ass:concave_game} and~\ref{ass:strongly_monotone_game}.
% \end{remark}

\begin{algorithm}[!t]
\caption{}\label{alg:bilevel_game}
\begin{algorithmic}[1]
    \State \textbf{Input:} $T$, $K$, $\epsilon\geq 0$, $\{\beta^t\}_{t=1}^{T}$. Set $\pi^1\in\Pi$ randomly.
    \For{\texttt{$t=1,\ldots, T$}}
        \State Leader announces $\pi^t$.
        \State \textbf{Inner loop:} Within $K$ rounds, followers compute \makebox[0.5cm][l]{}$\Tilde{x}(\pi^t)=\text{ApproxNE}(\pi^t, K)$ s.t.
        % \\ \hspace{2.8em} s.t. for some constant $C(\pi^t)$ and $\alpha>0$:
        \begin{align}\label{eq:approx_NE}
            \E[\lVert x^*(\pi^t)-\Tilde{x}(\pi^t)\rVert^2]\leq C(\pi^t)K^{-c}           
        \end{align}
        \State Followers set $x^t(\pi^t)=\Tilde{x}(\pi^t)$.
        \State Leader observes $J(\pi^t,x^t(\pi^t))$.
        %\State Leader appends $(\pi^t,x^t(\pi^t),\Tilde{J}(\pi^t,x^t(\pi^t);\pi^t))$ to history of play
        \State Leader updates $\tilde{\mu}^{t+1}$ and $\sigma^{t+1}$ via \eqref{eq:mu_corrected} and \eqref{eq:var}.
        \State Leader chooses:
        \begin{align*}\label{eq:next_pi}
            \pi^{t+1}\in\displaystyle\arg\min_{\pi\in\Pi} \til\mu^t(\pi)-\left(\beta^{t}+\frac{\epsilon\sqrt{t}}{\sigma}\right)\sigma^t(\pi)
        \end{align*}
    \EndFor
    %\State \textbf{Output:} \tcr{add}
\end{algorithmic}
\end{algorithm}

%%%%%%%% SECOND VERSION %%%%%%%%%%

We now provide our main result, which shows that the regret of the leader is sublinear in $T$ for properly chosen $\beta^t$. Furthermore, if the followers' utilities are Lipschitz, an approximate Stackelberg equilibrium is learned in time polynomial in $T$ and $K$.
\medskip

\begin{theorem}\label{th:1}
    Let Assumptions \ref{ass:concave_game} - \ref{ass:cost_in_RKHS} hold, set $\epsilon,\delta\in(0,1)$, and set $\beta^t$ equal to {\small$ B+(2L_JM/\sigma^2)\sqrt{2(\gamma^{t-1}+1+\log(2/\delta))}$}, where the \textit{maximum information gain} $\gamma^{t-1}$ is a kernel-dependent quantity defined in \cite{Srinivas2010}. 
    \begin{enumerate}
        \item With probability at least $1-\delta$ the following holds:
    \begin{align*}
        R^T\leq \mc O((1/\delta +\sqrt{\gamma^T}) TK^{-\frac{c}{2}}+\beta^T\sqrt{\gamma^TT})
    \end{align*}
      
    \noindent In other words, if
    $K=\mc O(T^{\frac{1}{c}})$ and $T=\mc O\left(\frac{1}{\epsilon^2}\right)$ with $c$ set as in Equation \eqref{eq:approx_NE} of Algorithm \ref{alg:bilevel_game},
    then with probability at least $1-\delta$ it holds that:
    %the sequence $\{\pi^t,x^t(\pi^t)\}_{t\geq 0}$ generated by Algorithm \ref{alg:bilevel_game} yields:
    %{
    %\begin{align*}
    %    R^T = \min_{\pi\in\Pi}\sum_{t=1}^T J(\pi^t, x^t(\pi^t))-J(\pi, x^*(\pi))\leq \epsilon.
    %\end{align*}
    %}
    \begin{align*}
    \min_{t\in[T]}J(\pi^{t},x^{t}(\pi^{t}))&\leq J(\pi^*,x^*(\pi^*)) + \epsilon.
    \end{align*}
    \item If in addition $U_i:\mc X\times\Pi$ is $L_{U_i}$-Lipschitz continuous in $x\in\mc X$ for all $\pi\in\Pi$ and all $i\in\mc N$, then with probability at least $1-\delta$ there exists a $t^*\in[T]$ such that $(\pi^{t^*},x^{t^*}(\pi^{t^*}))$  is an $\epsilon$-Stackelberg equilibrium.
    \end{enumerate}
\end{theorem}
\medskip
\noindent The regret bound in our theorem depends on two terms: 1) $\mc O(1/\delta L_j K^{-c/2}T)$, which stems from the error accumulation due to non-convergence of the inner-loop to the exact Nash equilibrium, and 2) $\mc O(\beta^T\sqrt{\gamma^T T})$, which stems from not knowing the true cost $J$ and estimating it using Gaussian process regression. Note that for some common kernels, explicit bounds on $\gamma^T$ are given, e.g., for the linear kernel $\gamma^T=\mc O(d\log T)$ and for the squared exponential kernel $\gamma^T=\mc O(\log^{d+1} T)$ \cite[Theorem~5]{Srinivas2010}, which depend sublinearly on $T$. By choosing $K$ large enough, i.e., $K=poly(T)$, the approximation error in the equilibrium-finding subroutine becomes negligible and our theorem shows that with high probability the leader finds an action that achieves nearly optimal cost function value in a sublinear number of rounds $T$. We now provide a proof of our theorem.
\medskip

\begin{proof}
We start by proving part 1) of the theorem. The regret of the leader is upper-bounded as follows:
{\begin{align}
        R^T &= \sum_{t=1}^T J(\pi^t, x^t(\pi^t)) - \min_{\pi\in\Pi}J(\pi, x^*(\pi)) \nonumber\\
        &\leq \sum_{t=1}^T \lvert J(\pi^t, x^t(\pi^t)) - J(\pi^t, x^*(\pi^t))\rvert \nonumber \\
        &+ \sum_{t=1}^T J(\pi^t, x^*(\pi^t)) - \min_{\pi\in\Pi}J(\pi, x^*(\pi)).
    \end{align}
    } 
    
    \noindent Then, due to Assumption \ref{ass:Lipschitz_cost}, we have:
{\begin{align}
        R^T &\leq  \underbrace{\sum_{t=1}^T L_J\lVert x^*(\pi^t) - x^t(\pi^t) \rVert}_{\coloneqq \Delta_F} \nonumber \\
        &+ \underbrace{ \sum_{t=1}^T J(\pi^t, x^*(\pi^t)) - \min_{\pi\in\Pi}J(\pi, x^*(\pi))}_{\coloneqq \Delta_L
        }.\label{eq:Rt}
    \end{align}
    }  
    
    \noindent We start by bounding the term $\Delta_F$. Note that by Jensen's inequality and Line \eqref{eq:approx_NE} in Algorithm \ref{alg:bilevel_game} the following holds:
    \begin{align}
        \bigl(\E\bigl[\lVert x^*(\pi^t) - x^t(\pi^t) \rVert\bigr]\bigr)^2&\leq\E\bigl[\lVert x^*(\pi^t) - x^t(\pi^t) \rVert^2\bigr] \nonumber\\
        \leq C(\pi^t)K^{-c}
        &\leq CK^{-c}\label{eq:Jensen},
    \end{align} 
    
    \noindent where in the last inequality we used that $\Pi$ is compact, which ensures the  existence of a maximizer $C=\max_{\pi\in\Pi}C(\pi)$. Furthermore, first applying Markov's inequality and then Inequality \eqref{eq:Jensen} yields:
    {
    \begin{align}
        \Prob(\Delta_F&\geq\frac{2}{\delta}TL_J  \sqrt{C}K^{-\frac{c}{2}}) \nonumber\\
        %&\leq\frac{\E\bigl[L_J\sum_{t=1}^T\lVert x^*(\pi^t)-x^t(\pi^t)\rVert\bigr]}{\frac{2L_J}{\delta}T_{\varepsilon}^{-\frac{\alpha}{2}}\sum_{t=1}^T\sqrt{C(\pi^t)}}\\
        &\leq\frac{\delta\sum_{t=1}^T\E\bigl[\lVert x^*(\pi^t)-x^t(\pi^t)\rVert \bigr]}{2 T\sqrt{C}K^{-\frac{c}{2}}} \leq \frac{\delta}{2}.\label{eq:prob_follower}
    \end{align}
    }
    
    \noindent Next, we bound the term $\Delta_L$, which corresponds to the regret $R^T$ of the leader assuming the lower-level game were at an exact Nash equilibrium $x^*(\pi^t)$.\footnote{Recall that at the exact Nash equilibrium $x^*(\pi^t)$ the leader's cost is uniquely defined by $J(\pi^t):=J(\pi^t,x^*(\pi^t))$.} Note that by Assumption \ref{ass:cost_in_RKHS} the cost function $J$ has a bounded RKHS norm. If the leader further observed $y^t=J(\pi^t,x^*(\pi^t))+\epsilon^t$, where $\epsilon^t$ is zero-mean and $R$-sub-Gaussian noise, then $\Delta_L$ could be upper-bounded leveraging the Gaussian process framework, for example by \cite[Theorem~3]{chowdhury2017kernelized}. In our setting, however, the zero-mean assumption on the noise is not necessarily satisfied. Namely, the leader observes $J(\pi^t,x^t(\pi^t))$ at each round $t$ rather than $J(\pi^t,x^*(\pi^t))$. The error term $\epsilon^t = J(\pi^t, x^*(\pi^t)) - J(\pi^t,x^t(\pi^t))$, resulting from approximating the Nash equilibrium, is not guaranteed to be zero-mean. 
    
    \noindent To alleviate this, we instead rewrite $\epsilon^t$ as:
    $$\epsilon^t = \E[J(\pi^t, x^*(\pi^t)) - J(\pi^t,x^t(\pi^t))] + \til \epsilon^t,$$ 
    where the expectation is taken with respect to the randomness in the inner-loop for finding $x^t(\pi^t)$, $\til\varepsilon^t$ is zero mean by construction, and $2L_JM$-sub-Gaussian as Assumption~\ref{ass:Lipschitz_cost} ensures the distribution of $\til\epsilon^t$ is bounded in $[-2L_J M, 2L_J M]$. 
    
    We furthermore define a new cost function $\til J:\Pi\rightarrow\R$ as $\til J(\pi,x^*(\pi))= J(\pi, x^*(\pi)) + \E[J(\pi,\Tilde{x}(\pi)) - J(\pi,x^*(\pi))]$, where $\Tilde{x}(\pi)=\text{ApproxNE}(\pi, K)$ in Algorithm \ref{alg:bilevel_game}, which corresponds to the true cost function at the Nash equilibrium plus the expected error from being at an approximation of the Nash equilibrium. Now the leader's observation $J(\pi^t,x^t(\pi^t))$ satisfies
    $J(\pi^t,x^t(\pi^t))=\til J(\pi^t,x^*(\pi^t))+\til \epsilon^t$, where $\til \epsilon^t$ is zero-mean, $2L_JM$-sub-Gaussian, but $\til J$ may no longer satisfy Assumption \ref{ass:cost_in_RKHS}, namely $\til J\not\in\{f\in\mc H_k(\Pi) \mid \lVert f\rVert_k\le B\}$. In the literature, this setting is referred to as the \emph{misspecified setting} \cite{bogunovic2021misspecified}. Note however that $\til J$ can be uniformly approximated by~$J$ as follows:
    \begin{align*}
    \lVert \til J - J\rVert_\infty&=\sup_{\pi\in\Pi} \lvert \til J(\pi,x^*(\pi))- J(\pi,x^*(\pi))\rvert\\
    &=\sup_{\pi\in\Pi} \lvert \E[ J(\pi,\til x(\pi))- J(\pi,x^*(\pi))]\rvert\\
    &\overset{(i)}{\le} L_J\sup_{\pi\in\Pi} \E[ \lVert x^*(\pi) -\til x(\pi)\rVert]\\
    &\overset{(ii)}{\le}L_J\sup_{\pi\in\Pi} \sqrt{\E[ \lVert x^*(\pi) -\til x(\pi)\rVert^2]}\\
    &\overset{(iii)}{\le} L_JC^{\frac{1}{2}}K^{-\frac{c}{2}},
\end{align*}
where in $(i)$ we used Assumption \ref{ass:Lipschitz_cost}, in $(ii)$ we applied Jensen's inequality, and in $(iii)$ we used Equation \eqref{eq:approx_NE}. In summary, the assumptions of \cite[Theorem~1]{bogunovic2021misspecified} are now satisified as $\til \epsilon^t$ is sub-Gaussian, $J\in\{f\in\mc H_k(\Pi) \mid \lVert f\rVert_k\le B\}$, and $\til J$ can be uniformly approximtaed by $J$. Then, since the leader chooses $\pi^t$ via Equation \eqref{eq:next_pi}, by \cite[Theorem~1]{bogunovic2021misspecified}, with probability at least $1-\delta/2$ it holds that:
% \begin{align}\label{eq:prob_leader}
% \Delta_L&=2\beta^T\sqrt{(2\sigma^2+1)\gamma^TT}+\\&+2\frac{L_JC^{\frac{1}{2}}K^{-\frac{c}{2}}}{\sigma}T\sqrt{(2\sigma^2+1)\gamma^T}.
% \end{align}
\begin{align}\label{eq:prob_leader}\Delta_L=2\sqrt{(2\sigma^2+1)\gamma^T}\big(\beta^T\sqrt{T}+T\frac{L_JC^{\frac{1}{2}}K^{-\frac{c}{2}}}{\sigma}\big)\end{align}
\noindent Next we bound $R^T$ leveraging the bounds we established for $\Delta_F$ and $\Delta_L$. To this end, let $E_F$ and $E_L$ denote the events:
    \begin{align*}
        E_F &= \big(\Delta_F\geq\frac{2 }{\delta}L_JTC^{\frac{1}{2}}K^{-\frac{c}{2}}\big)\\
        E_L &= \big(\Delta_L\geq 2\sqrt{(2\sigma^2+1)\gamma^T}\big(\beta^T\sqrt{T}+T\frac{L_JC^{\frac{1}{2}}K^{-\frac{c}{2}}}{\sigma}\big)
    \end{align*}
    and denote by $\ov E_F$ and $\ov E_L$ their corresponding complements. Then, it follows that:    
\begin{align*}
       &\Prob\big(R^T\leq \left(\frac{2 }{\delta}+\frac{2\sqrt{(2\sigma^2+1)\gamma^T}}{\sigma}\right)L_JTC^{\frac{1}{2}}K^{-\frac{c}{2}} + \\
        &\quad\quad 2\beta^T\sqrt{\gamma^TT(2\sigma^2+1)}\big) \\
        &\geq\Prob\big(\big(\Delta_F\leq \frac{2 }{\delta}L_JTC^{\frac{1}{2}}K^{-\frac{c}{2}}\big)\\
        &\quad\quad\cap \big(\Delta_L\leq   2\sqrt{(2\sigma^2+1)\gamma^T}(\beta^T\sqrt{T}+T\frac{L_J C^{\frac{1}{2}}K^{-\frac{c}{2}}}{\sigma}\big)\big)\\
        &= \Prob \left(\ov E_F \cap \ov E_L\right)\\
        &= 1-\Prob\bigl(E_F\cup E_L\bigr)\\
        &\geq 1- \bigl(\Prob(E_F)+\Prob(E_L)\bigr)\\
        &=1-(\frac{\delta}{2} + \frac{\delta}{2}).
    \end{align*}
    \noindent where we used the union bound in the second-to-last line and \eqref{eq:prob_follower} and \eqref{eq:prob_leader} in the last line. Finally, we have $\Prob \big(R^T\leq \mc O((\tfrac{1}{\delta}+\sqrt{\gamma^T})TK^{-\tfrac{c}{2}}+\beta^T\sqrt{T\gamma^T})\big)
        \geq 1-\delta$.

    Now, we proceed to prove part 2) of the theorem. Let $\pi^*\in\arg\min_{\pi\in\Pi}J(\pi,x^*(\pi))$. It can be easily verified that $(\pi^*, x^*(\pi^*))$ is a Stackelberg equilibrium. Set $t^*\in\arg\min_{t\in[T]} J(\pi^t,x^t(\pi^t))$, then, by definition of $R^T$ it follows that with probability at least $1-\delta$:
    \begin{align}\label{eq:bound_J}
        J(\pi^{t^*},x^{t^*}(\pi^{t^*}))&\leq J(\pi^*,x^*(\pi^*)) +2\frac{\beta^T}{\sqrt{T}}\sqrt{\gamma^T(2\sigma^2+1)}\big)\\
        &+\left(\frac{2 }{\delta}+\frac{2\sqrt{(2\sigma^2+1)\gamma^T}}{\sigma}\right)L_JC^{\frac{1}{2}}K^{-\frac{c}{2}}
    \end{align}
    Set $K$ and $T$ as follows: 
    {\small
    \begin{align}
        &K=\mc O(T^{\frac{1}{c}}) \label{eq:K}\\
        &T=\mc O\bigg(\frac{1}{\epsilon^2}\max\big\{ \displaystyle\max\{1, L_{U}^2\}\frac{L_J^2C}{\delta^2},\max\{ (\beta^T)^2,\frac{L_j^2C}{\sigma^2}\}\gamma^T\big\}\bigg) \label{eq:T}
        %& T \geq \max\{2;\quad\frac{4}{\epsilon^2} \frac{16}{\delta^2}L_j^2\max_{\pi} C(\pi);\quad    16(\beta^T)^2\gamma^T\frac{8}{\epsilon^2}\}
    \end{align}
    }
    \noindent where $L_U=\max_{i\in\mc N}L_{U_i}$. Then, by plugging the values of $K$ and $T$ into Equation \eqref{eq:bound_J} it follows that:
    {
    \begin{align*}
        J(\pi^{t^*},x^{t^*}(\pi^{t^*}))&\leq J(\pi^*,x^*(\pi^*)) + \epsilon\\
        &\leq J(\ov\pi, x^*(\ov\pi)) + \epsilon, \quad\forall\ov\pi\in\Pi,
    \end{align*}
    }
    
    \noindent where $x^*(\ov\pi)$ is the unique Nash equilibrium of the game {\small$\Gamma(\mc N, \{\mc X_i\}_{i=1}^N, \{U_i\}_{i=1}^N; \ov\pi)$}. Furthermore, if in addition $U_i:\mc X\times\Pi\rightarrow\R$ is $L_{U_i}$-Lipschitz continuous in $x\in\mc X$ for all $\pi\in\Pi$ and all $i\in\mc N$, then the following holds:
    
    {\small
    \begin{align*}
        &\Prob\bigg( \lvert U_i(x^*(\pi^{t^*}); \pi^{t^*}) - U_i(x^{t^*}(\pi^{t^*}); \pi^{t^*}) \lvert \geq \frac{\displaystyle L_{U_i}\sqrt{C} K^{-\frac{c}{2}}}{\delta }\bigg)\\
        &\leq \Prob\bigg( \lVert x^*(\pi^{t^*}) - x^{t^*}(\pi^{t^*}) \lVert \geq \frac{\sqrt{C} K^{-\frac{c}{2}}}{\delta}\bigg)\\
        &\leq \frac{\delta \E\big[\lVert x^*(\pi^{t^*}) - x^{t^*}(\pi^{t^*}) \rVert\big]}{\sqrt{C} K^{-\frac{c}{2}}} \leq \delta,
    \end{align*}
    }
    
    \noindent where in the last inequality we used Markov's inequality and Line \eqref{eq:approx_NE} in Algorithm \ref{alg:bilevel_game}.
    Thus, with $K$ and $T$ set as in Equations \eqref{eq:K} and \eqref{eq:T}, respectively, with probability at least $1-\delta$ it holds that:
    \begin{align*}
        U_i(x^{t^*}(\pi^{t^*}); \pi^{t^*}) +\epsilon &\geq U_i(x^*(\pi^{t^*}); \pi^{t^*})\\ 
        &\geq U_i(x_i,x_{-i}^*(\pi^{t^*}); \pi^{t^*}), \quad\forall x_i\in\mc X_i,
    \end{align*}
    where we used that $x^*(\pi^{t^*})$ is the unique Nash equilibrium of $\Gamma(\mc N, \{\mc X_i\}_{i=1}^N, \{U_i\}_{i=1}^N; \pi^{t^*})$. We conclude that $(\pi^{t^*},x^{t^*}(\pi^{t^*}))$ is an $\epsilon$
    -Stackelberg equilibrium.    
\end{proof}
\medskip

For a single-leader multi-follower Stackelberg game, \cite{maheshwari2023follower} shows that with $K=\mc O(T^\frac{1}{2c})$ and $T=\mc O(1/\epsilon^2)$, their algorithm converges to an $\epsilon$-stationary point, i.e., $\min_{t\in[T]}\E [\lVert \nabla J(\pi^t)\rVert^2]\leq \epsilon$. While their convergence time matches ours up to constant terms, their approach requires a hard-to-verify assumption on the Hessian of the leader's cost function to additionally ensure convergence to an approximate Stackelberg equilibrium. Moreover, their method requires two-point feedback to estimate the hypergradient of the leader's objective while our algorithm relies on one-point feedback to estimate the leader's cost function.

\definecolor{mycolor1}{RGB}{230,97,1}
\definecolor{mycolor2}{RGB}{178,171,210}
\begin {figure}%[!hbtp]
\centering
%\begin{adjustbox}{max height=0.4\textwidth, max width=\textwidth}
\resizebox{.45\textwidth}{!}{%
\begin{tikzpicture}[scale=0.75]

    \node (p1) at (0.5,0) {\includegraphics[width=.03\textwidth]{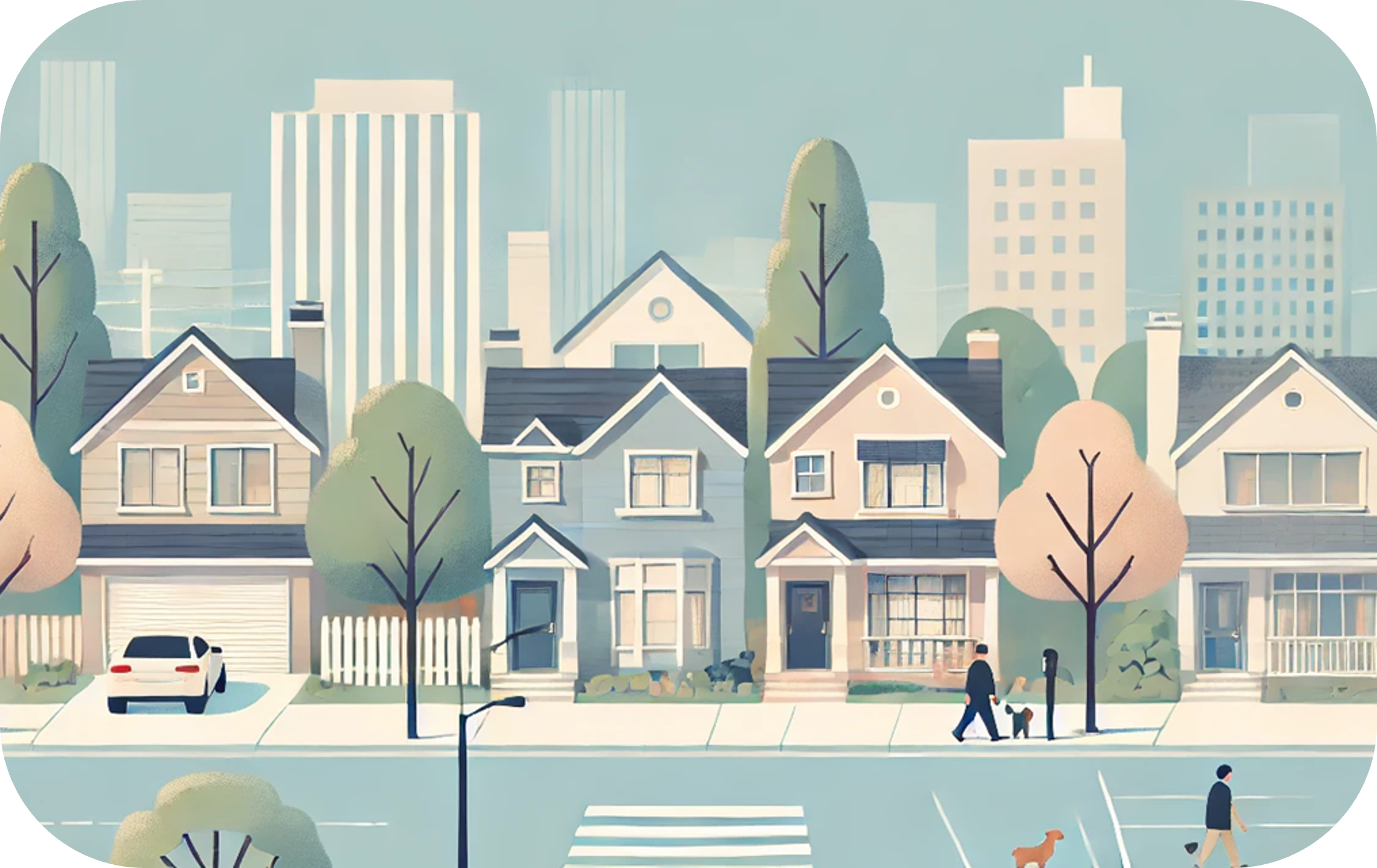}};
    \node[color=mycolor1, scale=0.12](dist1) at (0.5, 0.28) {Outskirts area: $W_1$, $\Delta_1$, $\pi_1$};
    
    \node (p2) at (0.5,-0.6) {\includegraphics[width=.03\textwidth]{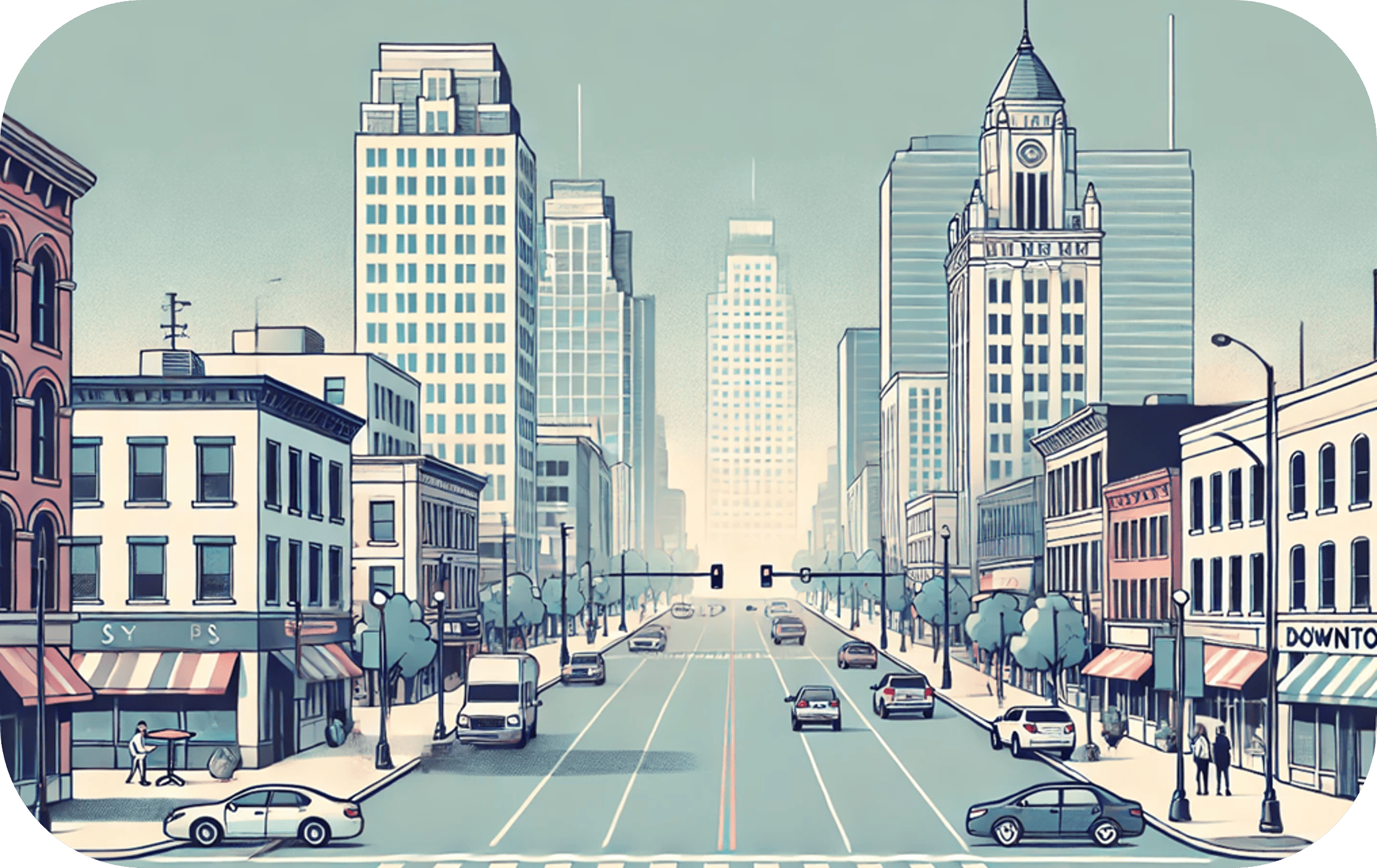}};
    \node[color=mycolor2, scale=0.12](dist1) at (0.5, -0.32) {Downtown area: $W_2$, $\Delta_2$, $\pi_2$};
    
    %\draw[rounded corners=1pt, line width=0.05pt] (0.02,0.35) rectangle (0.98,-0.9);

    \node (p3) at (-0.505,0.19) {\includegraphics[width=.015\textwidth]{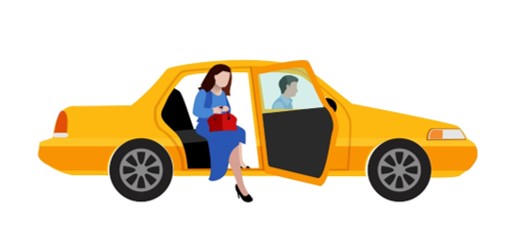}};
    \node (p4) at (-0.77,0.23) {\includegraphics[width=.01\textwidth]{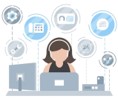}};
    \node[ scale=0.1](n3) at (-0.5, 0.28) {$M_1$ vehicles};
    \draw[rounded corners=1pt, line width=0.05pt] (-0.9,0.33) rectangle (-0.35, 0.12);
    \node[ scale=0.1](n33) at (-0.625, 0.36) {\textbf{Company} 1};
    
    \node (p5) at (-0.505,-0.31) {\includegraphics[width=.015\textwidth]{figures/taxi.jpg}};
    \node (p6) at (-0.77,-0.27) {\includegraphics[width=.01\textwidth]{figures/Operator.jpg}};
    \node[ scale=0.1](n4) at (-0.5, -0.22) {$M_2$ vehicles};
    \draw[rounded corners=1pt, line width=0.05pt] (-0.9,-0.17) rectangle (-0.35, -0.38);
    \node[ scale=0.1](n33) at (-0.625, -0.14) {\textbf{Company} 2};

    \node (p7) at (-0.505,-0.81) {\includegraphics[width=.015\textwidth]{figures/taxi.jpg}};
    \node (p8) at (-0.77,-0.77) {\includegraphics[width=.01\textwidth]{figures/Operator.jpg}};
    \node[ scale=0.1](n5) at (-0.5, -0.72) {$M_3$ vehicles};
    \draw[rounded corners=1pt, line width=0.05pt] (-0.9,-0.67) rectangle (-0.35, -0.88);
    \node[ scale=0.1](n33) at (-0.625, -0.64) {\textbf{Company} 3};

    %\draw[-{Triangle[length=0.8pt, width=0.5pt]}, line width=0.05pt] (-0.35,0.225) -- (0.02, -0.275);
    %\draw[-{Triangle[length=0.8pt, width=0.5pt]}, line width=0.05pt] (-0.35,-0.275) -- (0.02, -0.275);
    %\draw[-{Triangle[length=0.8pt, width=0.5pt]}, line width=0.05pt] (-0.35,-0.775) -- (0.02, -0.275);
    
    \draw[-{Triangle[length=0.8pt, width=0.5pt]}, line width=0.05pt] (-0.35,0.225) -- (0.14, 0.0);
    \draw[-{Triangle[length=0.8pt, width=0.5pt]}, line width=0.05pt] (-0.35,0.225) -- (0.14, -0.6);    
    \draw[-{Triangle[length=0.8pt, width=0.5pt]}, line width=0.05pt] (-0.35,-0.275) -- (0.14, 0.0);
    \draw[-{Triangle[length=0.8pt, width=0.5pt]}, line width=0.05pt] (-0.35,-0.275) -- (0.14, -0.6);
    \draw[-{Triangle[length=0.8pt, width=0.5pt]}, line width=0.05pt] (-0.35,-0.775) -- (0.14, 0.0);
    \draw[-{Triangle[length=0.8pt, width=0.5pt]}, line width=0.05pt] (-0.35,-0.775) -- (0.14, -0.6);

    \node[color=mycolor1, scale=0.11](x11) at (-0.28,0.25){$x_{1,1}$};
    \node[color=mycolor2,scale=0.11](x12) at (-0.28,0.00){$x_{1,2}$};

    \node[color=mycolor1,scale=0.11](x21) at (-0.28,-0.17){$x_{2,1}$};
    \node[color=mycolor2,scale=0.11](x22) at (-0.28,-0.38){$x_{2,2}$};

    \node[color=mycolor1,scale=0.11](x31) at (-0.28,-0.57){$x_{3,1}$};
    \node[color=mycolor2,scale=0.11](x32) at (-0.28,-0.8){$x_{3,2}$};
    
    \draw[rounded corners=1pt, line width=0.05pt] (-0.95, 0.4) rectangle (1.03, -0.95);

    \node (p10) at (-0.16, 0.725) {\includegraphics[width=.002\textwidth]{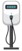}};
    \draw[fill, color=mycolor1, line width=0.05pt] (-0.11, 0.75) rectangle (-0.04, 0.68);
    \node[color=mycolor1,scale=0.1](n6) at (-0.11, 0.63) {Station 1};

    \node (p11) at (0.14, 0.725) {\includegraphics[width=.002\textwidth]{figures/charger.jpg}};
    \draw[fill, color=mycolor2, line width=0.05pt] (0.19, 0.75) rectangle (0.26, 0.68);
    \node[color=mycolor2,scale=0.1](n6) at (0.19, 0.63) {Station 2};

    \draw[rounded corners=1pt, line width=0.05pt] (-0.56, 0.8) rectangle (0.64, 0.6);

    \node (p13) at (-0.135, 1.3) {\includegraphics[width=.01\textwidth]{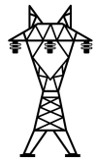}};
    \draw[rounded corners=1pt, line width=0.05pt] (-0.435, 1.53) rectangle (0.165, 1.07);
    \draw[rounded corners=1pt, line width=0.05pt, fill=white] (-0.085, 1.4) rectangle (0.515, 0.94);
    \node (p14) at (0.215, 1.17) {\includegraphics[width=.023\textwidth]{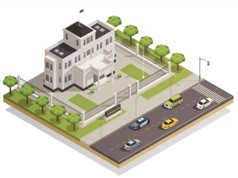}};
    \draw[rounded corners=1pt, line width=0.05pt] (-0.475, 1.57) rectangle (0.555, 0.9);
    \node[scale=0.13](ca) at (0.04,1.6){\textbf{Central authority}};

    \node[scale=0.1](ci) at (0.04, 0.85){\textbf{Charging infrastructure}};
    \draw[-{Triangle[length=0.8pt, width=0.5pt]}, line width=0.05pt] (0.04, 0.6) -- (0.04, 0.4);

    \node[scale=0.13](pr) at (0.305,0.5){Prices $\pi=(\textcolor{mycolor1}{\pi_1}, \textcolor{mycolor2}{\pi_2})$};
    \node[scale=0.13](rh) at (-0.625,0.43){\textbf{Ride-hailing market}};

\end{tikzpicture}}%
%\end{adjustbox}
    \caption{Illustration of the setup with 2 districts and 3 ride-hailing companies.} \vspace{-1.5em}
\label{fig:market}
\end{figure}
% \begin {figure}%[!hbtp]
% \centering
% \begin{adjustbox}{max height=0.55\textwidth, max width=0.48\textwidth}
% \begin{tikzpicture}[scale=1.0, font=\Large]

%     \node[scale=3] (pic) at (0,0) {\input{figures/Setup.tikz}};

% \end{tikzpicture}
% \end{adjustbox}
%     \caption{Illustration of the setup with 2 districts and 3 ride-hailing companies.} \vspace{-1.5em}
% \label{fig:market}
% \end{figure}

\section{Coordination of ride-hailing companies: Numerical study}\label{sec:numerical}

We apply Algorithm \ref{alg:bilevel_game} to the pricing problem in electric ride-hailing markets, introduced in Section~\ref{sec:mot_examp}. In our numerical study illustrated in Figure~\ref{fig:market}, for the lower level, we consider a city region divided into two districts, the outskirts, and a downtown area, where $|\mc N|=3$ competing ride-hailing companies operate and can recharge their fleets of scaled sizes $M=(2, 4, 6)$. To reflect the typical higher demand in downtown areas, we set the total revenue potential and abandonment vectors in Equation~\eqref{eq:Player_util} to $W=(30, 60)$ and $\Delta=(0.1, 0.5)$, respectively. 

In this simplified setup, we assume that the regulatory authority aims to select an optimal charging price vector $\pi\in\R^2$ within the range $[0.1, 5.0]$ to help balance the demand on the power grid and reduce the number of idling vehicles in the downtown area. Specifically, to counterbalance the increased attractiveness of the downtown district, the authority aims to set lower charging prices in the outskirts, with the goal of steering the vehicle distribution toward a uniform spread, i.e., $\xi^*=(0.5, 0.5)$. For example, if the central authority sets prices proportional to the revenue potential of the districts, e.g., $\pi_{\text{base}} = (1, 2)$, then the attained leader’s cost equals $J(\pi_{\text{base}}) \approx 0.22$ while $J(\pi^*)=0$ if the central authority sets prices optimally.

We implement Algorithm \ref{alg:bilevel_game} by choosing a squared exponential kernel, combined with a standard heuristic approach from~\cite{brochu2009tutorial} to approximate the minimum of the optimistic estimate of the cost function in Line 8 of Algorithm~\ref{alg:bilevel_game}. To demonstrate the practicality of our learning method, we evaluate its performance under various levels of approximation error when solving the lower-level Nash equilibrium. Specifically, we terminate the inner loop when $\norm{x^*(\pi^t)-x^t(\pi^t)} \leq \varepsilon$ for $\varepsilon \in \{10^{-6}, 0.1, 0.3, 0.5\}$. For each $\varepsilon$, the GP parameters are calibrated via standard gradient-based optimization of the marginal log-likelihood~\cite{rasmussen2005}, using data collected after $N_{\text{warm}} = 5$ random iterations of the outer loop. While the value of $\beta^T$ proposed by Theorem~\ref{th:1} provides theoretical guarantees, in practice we empirically found that fixing $\beta^T=0.2$ enables the leader to find a high-quality pricing for all $\varepsilon$ values, while avoiding the computational burden of finding the problem-specific constants required by Theorem~\ref{th:1}. 

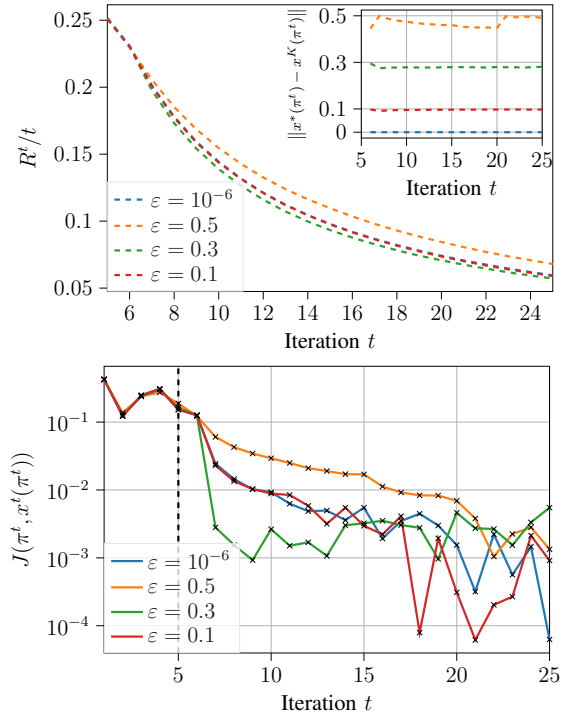
\begin {figure}%[!hbtp]
\centering
\begin{adjustbox}{max width=0.45\textwidth}
\begin{tikzpicture}[scale=1.0]

    \node[scale=0.7] (pic1) at (0.0, 0.0) {\begin{tikzpicture}

\definecolor{crimson2143940}{RGB}{214,39,40}
\definecolor{darkgrey176}{RGB}{176,176,176}
\definecolor{darkorange25512714}{RGB}{255,127,14}
\definecolor{forestgreen4416044}{RGB}{44,160,44}
\definecolor{steelblue31119180}{RGB}{31,119,180}

\begin{axis}[
font=\large, % Bigger fonts
legend cell align={left},
legend style={font=\large, fill opacity=0.8, draw opacity=1, text opacity=0.9, draw=white!90!black, at={(0.00,0.00)}, anchor=south west},
name=ax1,
ytick={0.05, 0.1, 0.15, 0.2, 0.25},
yticklabels={%
$0.05$,
$0.1$,
$0.15$,
$0.2$,
$0.25$,
},
tick align=outside,
tick pos=left,
x grid style={darkgrey176},
xlabel={Iteration $t$},
xmin=5, xmax=25,
xtick style={color=black},
y grid style={darkgrey176},
ylabel={$R^t/t$},
ymin=0.0474768634482884, ymax=0.261530065326198,
ytick style={color=black},
width=10cm,
height=7cm,
]

\addplot [very thick, steelblue31119180, dashed]
table {%
5 0.251800374331748
6 0.230883484384219
7 0.201379938548197
8 0.178015175284714
9 0.159372819019998
10 0.144362696395021
11 0.131809300523157
12 0.121231124572694
13 0.112288022032965
14 0.104524043924515
15 0.0979213904448637
16 0.0919217307154284
17 0.086719948653199
18 0.0821504720863748
19 0.0779841976858406
20 0.07416209336202
21 0.0706457104785078
22 0.0675356154189665
23 0.0646238371809807
24 0.061991930735296
25 0.0595147686598727
};
\addlegendentry{$\varepsilon=10^{-6}$}

\addplot [very thick, darkorange25512714, dashed]
table {%
5 0.25122880365373
6 0.229809060379189
7 0.205602146033374
8 0.18524576916153
9 0.168487850254076
10 0.154571749538229
11 0.142788563442176
12 0.132633768041838
13 0.12389516211191
14 0.116267209413275
15 0.109646463763293
16 0.103495744222856
17 0.0979483938531066
18 0.0929699119676527
19 0.0885119437851909
20 0.0844325394787574
21 0.0805931004228864
22 0.0769773142583978
23 0.0737281359823601
24 0.0707735774085687
25 0.0679961103542676
};
\addlegendentry{$\varepsilon=0.5$}

\addplot [very thick, forestgreen4416044, dashed]
table {%
5 0.251222014760128
6 0.230179339278058
7 0.197698066753728
8 0.173178869677434
9 0.154039803981284
10 0.138899566621404
11 0.126409352954978
12 0.11601631402866
13 0.107174306256094
14 0.0997344300902173
15 0.0933007869106887
16 0.08768940200383
17 0.0827100626766119
18 0.0782690931012481
19 0.0742005549784548
20 0.0707213399865781
21 0.067483387284127
22 0.0645368156681074
23 0.0617977187977539
24 0.0593616682333951
25 0.0572065544427388
};
\addlegendentry{$\varepsilon=0.3$}

\addplot [very thick, crimson2143940, dashed]
table {%
5 0.251059129169153
6 0.230250229688945
7 0.200639473876742
8 0.177239009969182
9 0.158690184261679
10 0.143707752437101
11 0.131409979466916
12 0.120945251908164
13 0.111885787685924
14 0.1042858693378
15 0.0975311415463158
16 0.091573606569966
17 0.0864279204853961
18 0.0816307633063538
19 0.0774363414963254
20 0.073580034477998
21 0.0700791634228452
22 0.066902988559049
23 0.0640058091711125
24 0.0614274748319584
25 0.0590068880342203
};
\addlegendentry{$\varepsilon=0.1$}

\end{axis}

\begin{axis}[
font=\large, % Bigger font for inset
xtick={0, 5, 10, 15, 20, 25},
xticklabels={%
$0$,
$5$,
$10$,
$15$,
$20$,
$25$,
},
ytick={0, 0.1, 0.3, 0.5},
yticklabels={%
$0$,
$0.1$,
$0.3$,
$0.5$,
},
name=ax2,
tick align=outside,
tick pos=left,
width=5cm,
height=4cm,
x grid style={darkgrey176},
xlabel={Iteration $t$},
xmajorgrids,
xmin=5, xmax=25,
xtick style={color=black},
y grid style={darkgrey176},
ylabel={\footnotesize$\norm{x^*(\pi^t)-x^K(\pi^t)}$},
ymajorgrids,
ymin=-0.0249817656285801, ymax=0.524617078200183,
ytick style={color=black}, 
at={($(ax1.south west)+(4.8cm,2.9cm)$)},
]

\addplot [very thick, steelblue31119180, dashed]
table {%
6 0
7 0
8 0
9 0
10 0
11 0
12 0
13 0
14 0
15 0
16 0
17 0
18 0
19 0
20 0
21 0
22 0
23 0
24 0
25 0
};

\addplot [very thick, darkorange25512714, dashed]
table {%
6 0.4450224267175
7 0.499635312571603
8 0.486791490560079
9 0.479217734484287
10 0.474119449383337
11 0.46947514307571
12 0.464976985913962
13 0.46278271305763
14 0.460508115514861
15 0.460331812986688
16 0.453297618191053
17 0.450686237282216
18 0.449578795386597
19 0.449491380894526
20 0.447730849571584
21 0.498041373391299
22 0.493723414078584
23 0.495614093427252
24 0.496509509308244
25 0.489917860378978
};

\addplot [very thick, forestgreen4416044, dashed]
table {%
6 0.295809235779487
7 0.274365901174084
8 0.277922682946182
9 0.277429156917333
10 0.278787956830779
11 0.277892978366369
12 0.278040816253928
13 0.277543760968535
14 0.279085112359902
15 0.279252280579692
16 0.279477510159272
17 0.279104368651545
18 0.278894207315257
19 0.277460919074344
20 0.280326272908735
21 0.27885628137957
22 0.278804773738182
23 0.277917183124364
24 0.27933246864278
25 0.280989131631343
};

\addplot [very thick, crimson2143940, dashed]
table {%
6 0.0974229707450033
7 0.0912166140468119
8 0.0940558797588981
9 0.0949015570456404
10 0.0952732762616928
11 0.0953838436426273
12 0.0960283457845401
13 0.0966555697202761
14 0.0961118550077457
15 0.0967030238278925
16 0.0968740226114327
17 0.0964413785097111
18 0.0973425511812027
19 0.096935426482063
20 0.0972927991785742
21 0.097346275635555
22 0.0973158405059854
23 0.0973019327065634
24 0.096893075402873
25 0.0971618720319525
};

\end{axis}

\end{tikzpicture}
};

    \node[scale=0.7] (pic2) at (0.0, -4.8) {% This file was created with tikzplotlib v0.10.1.
\begin{tikzpicture}

\definecolor{crimson2143940}{RGB}{214,39,40}
\definecolor{darkgrey176}{RGB}{176,176,176}
\definecolor{darkorange25512714}{RGB}{255,127,14}
\definecolor{forestgreen4416044}{RGB}{44,160,44}
\definecolor{lightgrey204}{RGB}{204,204,204}
\definecolor{steelblue31119180}{RGB}{31,119,180}

\begin{axis}[
font=\large, % Bigger fonts
legend cell align={left},
legend style={font=\large, fill opacity=0.8, draw opacity=1, text opacity=0.9, draw=white!90!black, at={(0.00,0.00)}, anchor=south west},
log basis y={10},
tick align=outside,
tick pos=left,
x grid style={darkgrey176},
xlabel={Iteration $t$},
xmajorgrids,
xmin=1, xmax=25,
xtick style={color=black},
y grid style={darkgrey176},
ylabel={$J(\pi^t,x^t(\pi^t))$},
ymajorgrids,
ymin=3.96839943023755e-05, ymax=0.663592992221667,
ymode=log,
ytick style={color=black},
ytick={1e-06,1e-05,0.0001,0.001,0.01,0.1,1,10},
yticklabels={
  \(\displaystyle {10^{-6}}\),
  \(\displaystyle {10^{-5}}\),
  \(\displaystyle {10^{-4}}\),
  \(\displaystyle {10^{-3}}\),
  \(\displaystyle {10^{-2}}\),
  \(\displaystyle {10^{-1}}\),
  \(\displaystyle {10^{0}}\),
  \(\displaystyle {10^{1}}\)
},
width=10cm,
height=7cm,
]

% Data points (kept thin so markers stand out)
\addplot [draw=black, fill=black, forget plot, mark=x, only marks]
table{%
x  y
1 0.426514918980194
2 0.122541081583682
3 0.249093389557237
4 0.308759358886161
5 0.152093122651464
6 0.126299034646575
7 0.024358663532066
8 0.0144618324403343
9 0.0102339689022683
10 0.00927159277023306
11 0.00627534180450799
12 0.00487118911760006
13 0.00497079155622829
14 0.00359232851465552
15 0.00548424172975002
16 0.00192683477389782
17 0.00349143565752964
18 0.00446937045036292
19 0.0029912584762239
20 0.00154211120942904
21 0.000318052808264063
22 0.00222361916859954
23 0.000564715945293533
24 0.00145808248454722
25 6.28788497129483e-05
};

\addplot [draw=black, fill=black, forget plot, mark=x, only marks]
table{%
x  y
1 0.418554441833428
2 0.13737999332861
3 0.239382559214432
4 0.273156233964669
5 0.187670789927512
6 0.122710344006485
7 0.06036065995848
8 0.0427511310586254
9 0.0344244989944432
10 0.0293268430956038
11 0.0249567024816519
12 0.0209310186381217
13 0.0190318909527654
14 0.0171038243310226
15 0.0169560246635419
16 0.011234951116303
17 0.00919078793711759
18 0.00833571991493764
19 0.00826851650087718
20 0.00692385765652152
21 0.00380431930546595
22 0.00104580480413885
23 0.0022462139095302
24 0.00281873021136556
25 0.00133690105104078
};

\addplot [draw=black, fill=black, forget plot, mark=x, only marks]
table{%
x  y
1 0.424966332616325
2 0.126727586370779
3 0.239382559214432
4 0.301125955351773
5 0.163907640247334
6 0.124965961867703
7 0.00281043160775099
8 0.00154449014337749
9 0.000927278412085469
10 0.00263743038248388
11 0.00150721629071159
12 0.00169288583916052
13 0.00107021298531115
14 0.00301603993381568
15 0.00322978239728817
16 0.00351862840094866
17 0.00304063344112304
18 0.00277261032006308
19 0.000966868768175327
20 0.00461625514092071
21 0.00272433323510546
22 0.00265881173169552
23 0.00153758764997685
24 0.00333250525314165
25 0.00548382346698843
};

\addplot [draw=black, fill=black, forget plot, mark=x, only marks]
table{%
x  y
1 0.426389312094807
2 0.122553241650256
3 0.245419295228087
4 0.308062559389537
5 0.152871237483078
6 0.126205732287905
7 0.0229749390035276
8 0.0134357626162611
9 0.0102995786016572
10 0.00886586601589709
11 0.00843224976506252
12 0.00583324876189276
13 0.0031722170190428
14 0.00548693081219178
15 0.00296495246553534
16 0.00221058192471774
17 0.00409694313227836
18 7.90912626352556e-05
19 0.00193674891581465
20 0.000310201129777183
21 6.17423197889212e-05
22 0.000203316419328675
23 0.000267862636508363
24 0.00212578503141501
25 0.000912804888505672
};

% Main lines (made very thick)
\addplot [very thick, steelblue31119180]
table {%
1 0.426514918980194
2 0.122541081583682
3 0.249093389557237
4 0.308759358886161
5 0.152093122651464
6 0.126299034646575
7 0.024358663532066
8 0.0144618324403343
9 0.0102339689022683
10 0.00927159277023306
11 0.00627534180450799
12 0.00487118911760006
13 0.00497079155622829
14 0.00359232851465552
15 0.00548424172975002
16 0.00192683477389782
17 0.00349143565752964
18 0.00446937045036292
19 0.0029912584762239
20 0.00154211120942904
21 0.000318052808264063
22 0.00222361916859954
23 0.000564715945293533
24 0.00145808248454722
25 6.28788497129483e-05
};
\addlegendentry{$\varepsilon=10^{-6}$}

\addplot [very thick, darkorange25512714]
table {%
1 0.418554441833428
2 0.13737999332861
3 0.239382559214432
4 0.273156233964669
5 0.187670789927512
6 0.122710344006485
7 0.06036065995848
8 0.0427511310586254
9 0.0344244989944432
10 0.0293268430956038
11 0.0249567024816519
12 0.0209310186381217
13 0.0190318909527654
14 0.0171038243310226
15 0.0169560246635419
16 0.011234951116303
17 0.00919078793711759
18 0.00833571991493764
19 0.00826851650087718
20 0.00692385765652152
21 0.00380431930546595
22 0.00104580480413885
23 0.0022462139095302
24 0.00281873021136556
25 0.00133690105104078
};
\addlegendentry{$\varepsilon=0.5$}

\addplot [very thick, forestgreen4416044]
table {%
1 0.424966332616325
2 0.126727586370779
3 0.239382559214432
4 0.301125955351773
5 0.163907640247334
6 0.124965961867703
7 0.00281043160775099
8 0.00154449014337749
9 0.000927278412085469
10 0.00263743038248388
11 0.00150721629071159
12 0.00169288583916052
13 0.00107021298531115
14 0.00301603993381568
15 0.00322978239728817
16 0.00351862840094866
17 0.00304063344112304
18 0.00277261032006308
19 0.000966868768175327
20 0.00461625514092071
21 0.00272433323510546
22 0.00265881173169552
23 0.00153758764997685
24 0.00333250525314165
25 0.00548382346698843
};
\addlegendentry{$\varepsilon=0.3$}

\addplot [very thick, crimson2143940]
table {%
1 0.426389312094807
2 0.122553241650256
3 0.245419295228087
4 0.308062559389537
5 0.152871237483078
6 0.126205732287905
7 0.0229749390035276
8 0.0134357626162611
9 0.0102995786016572
10 0.00886586601589709
11 0.00843224976506252
12 0.00583324876189276
13 0.0031722170190428
14 0.00548693081219178
15 0.00296495246553534
16 0.00221058192471774
17 0.00409694313227836
18 7.90912626352556e-05
19 0.00193674891581465
20 0.000310201129777183
21 6.17423197889212e-05
22 0.000203316419328675
23 0.000267862636508363
24 0.00212578503141501
25 0.000912804888505672
};
\addlegendentry{$\varepsilon=0.1$}

% Vertical dashed line
\addplot [very thick, black, dashed, forget plot]
table {%
5 3.96839943023755e-05
5 0.663592992221667
};

\end{axis}

\end{tikzpicture}
};
    
\end{tikzpicture}
\end{adjustbox}
    \caption{The top figure illustrates the average cumulative regret of the regulatory authority, $R^t/t$, while the bottom figure displays the leader’s objective, both over $T = 25$ iterations. The initial $N_{\text{warm}} = 5$ iterations correspond to a warm-up phase, during which pricing vectors are selected randomly in order to collect data for calibrating hyperparameters of the GP. Different colors represent varying levels of approximation error in computing the Nash equilibrium within the inner loop of Algorithm~\ref{alg:bilevel_game}.} \vspace{-1.5em}
\label{fig:res}
\end{figure}

The results are illustrated in Figure~\ref{fig:res} and further supported by numerical values in Table~\ref{tab:1}.  
\begin{table}%[!hbtp]
\begin{center}
 \renewcommand{\arraystretch}{1.2}
 \caption{Charging prices and leader's performance}\label{tab:1}
 \begin{tabular}{c|cc|c|c}
 %\hline
 \multirow{2}{*}{\makecell{Approximation \\ error $\varepsilon$} }  & \multicolumn{2}{c|}{Best pricing vector} & \multirow{2}{*}{$J(\pi^t, x^t(\pi^t))$ } &  \multirow{2}{*}{$R^t/t$ }\\ %\cline{2-9}
 &$\pi_1^t$ & $\pi_2^t$ & &  \\
 \hline
 \hline
 $10^{-6}$ &  1.9157 & 4.99 & $6.3\cdot 10^{-5}$  & 0.0595   \\
 $0.1$     &  1.9134 & 4.99 & $6.2\cdot 10^{-5}$  & 0.0590   \\
 $0.3$     &  1.8761 & 4.99 & $9.2\cdot 10^{-4}$  & 0.0572  \\
 $0.5$     &  1.8305 & 4.99 & $1.3\cdot 10^{-3}$  & 0.0679   \\ 
\hline 
\end{tabular}
\end{center}
\vspace{-1.5em}
\end{table}
The cumulative regret plot suggests that our proposed framework is fairly robust to approximation errors in the inner loop, as all curves for $\varepsilon \leq 0.3$ show a similar downward trend over time, with the $\varepsilon=10^{-6}$ and $\varepsilon=0.1$ curves almost perfectly overlapping. For $\varepsilon = 0.5$, the trend remains consistent but decreases at a slightly slower rate. Interestingly, the heuristic  from~\cite{brochu2009tutorial}, when combined with a fixed $\beta^T$, led to the fastest reduction in average cumulative regret and the most rapid initial learning for $\varepsilon = 0.3$. However, the lowest leader’s objective is obtained when the approximation error is negligible.

\section{Conclusion}\label{sec:conclusion}

%In this work, we tackled the challenge of computing Stackelberg equilibria in multi-agent systems, with the aim of learning effective leader incentives. Motivated by electric ride-hailing markets, we proposed BIG GP-LCB, a novel no-regret algorithm that enables the leader to solve Stackelberg games with minimal feedback. By treating the lower-level game as a black box and assuming repeated follower interactions to approximate Nash equilibria, we established high-probability guarantees on the convergence of BIG GP-LCB to an $\epsilon$-Stackelberg equilibrium in $\mathcal{O}(\sqrt{T})$ rounds. 

In this paper, we studied the problem of learning to play single-leader, multi-follower Stackelberg games when the lower-level game is unknown to the leader. We proposed a novel no-regret algorithm, and proved that it converges to an $\epsilon$-Stackelberg equilibrium in $\mc O(\sqrt{T})$ rounds with high probability under kernel-based regularity assumptions. Our method improves practicality by removing the need for gradient-based techniques that require access to followers' private utilities. Lastly, we validated our method in a numerical case study on electricity pricing, demonstrating its convergence under varying levels of lower-level approximation error. Future work may explore extensions to contextual settings, where the lower-level game depends on both the leader's action and a random context, accounting for external factors like weather and time that are beyond the leader's control.% where a random context beyond the leader's control  and real-world deployment in broader socio-technical systems.

%\section*{ACKNOWLEDGMENT}

%%%%%%%%%%%%%%%%%%%%%%%%%%%%%%%%%%%%%%%%%%%%%%%%%%%%%%%%%%%%%%%%%%%%%%%%%%%%%%%%

\bibliographystyle{IEEEtran}
\bibliography{references.bib}

\end{document}